\documentclass[11pt]{article}

\usepackage{epsfig,graphicx}
\usepackage{amsbsy}
\usepackage{amsfonts}

\newcommand{\bea}{\begin{eqnarray}}
\newcommand{\eea}{\end{eqnarray}}

\begin{document}

\pagestyle{empty}  
  
\begin{flushright}   
  
{\tt hep-ph/0607199}\end{flushright}   
  
\begin{center}   
\vspace*{0.5cm}  
{\Large \bf Neural network determination of the non-singlet quark distribution}\\  
\vspace*{1.5cm}   
{\bf The NNPDF Collaboration:} \\
\vspace{0.6cm}  
Andrea Piccione$^1$, \\
Luigi Del Debbio$^2$, Stefano
  Forte$^3$, Jos\'e I. Latorre$^4$ and 
Joan Rojo$^4$
\vspace{0.6cm}  

~$^1$ Dipartimento di Fisica Teorica, Universit\`a di Torino and 
INFN, Sezione di Torino,  \\Via P.~Giuria 1, I-10125 Torino, Italy\\
\vspace{0.2cm}  
~$^2$ Schhol of Physics, University  of Edinburgh\\
 Edinburgh EH9 3JZ Scotland\\
\vspace{0.2cm}  
~$^3$ Dipartimento di Fisica, Universit\`a di Milano and
INFN, Sezione di Milano, \\Via Celoria 16, I-20133 Milano, Italy\\
\vspace{0.2cm}  
~$^4$ Departament d'Estructura i Constituents de la Mat\`eria, \\
Universitat de Barcelona, Diagonal 647, E-08028 Barcelona, Spain\\

\vspace*{0.6cm}  
  
{\bf Abstract}  
\end{center}  

We summarize the main features of our approach to parton fitting, and
we show a preliminary result for the non-singlet structure
function. When comparing our result to other PDF sets, we find a
better description of large $x$ data and larger error bands in the
extrapolation regions.

\noindent  
\vspace*{1cm}  
\vfill  
\noindent  
  
\begin{flushleft} July 2006 \end{flushleft}   
\eject   
  
\setcounter{page}{1} \pagestyle{plain}  

\section{The NNPDF approach}
\newcommand{\xuno}{\frac{\omega_{11}^{(2)}}
{1+e^{\theta_1^{(2)}-\xi_1^{(1)}\omega_{11}^{(1)}}}}
\newcommand{\xdue}{\frac{\omega_{12}^{(2)}}
{1+e^{\theta_2^{(2)}-\xi_1^{(1)}\omega_{21}^{(1)}}}}

The standard approach to PDF fitting has two main shortcomings. The first
is the difficulty in propagating the error from data to the
parametrization, and then from the parametrization to any observable
that it is evaluated with it: this is easy to do only in a linearized
approximation, which is not always adequate. The second is the
difficulty in assessing the bias associated to the choice of
functional form, which is done on the basis of theoretical
prejudice. The latter is especially delicate, because a functional
form parametrized by a small number of parameters must be chosen in
order for the fits to converge, but this is then inevitably a source
of bias: a bias free fit would never converge.

We have proposed a new approach to this problem
\cite{Forte:2002fg,DelDebbio:2004qj}, which is based on the use of
neural networks combined with the Monte Carlo method. The
Monte Carlo approach addresses the first difficulty of the standard
approach. Instead of propagating the experimental error on the
parameters of the parton distributions, we generate replicas of the
true experimental data, which fluctuate about the central experimental
values in a way that reproduces the data uncertainty. If the number of
replicas is sufficiently large, averaging over the replicas we can
reconstruct the data we started from with their errors and
correlations. Instead of producing a single set of parton
distributions, we then produce as many replicas of the parton
distributions as we generated replicas of the original data. The
fluctuation of these replicas then automatically propagates the
fluctuations of the data we started from, and averaging over them we
can reconstruct the value and uncertainty on the parton distributions,
and indeed of any physical observable which depends on them.

In order to avoid any assumption on the shape of the PDF at the
initial scale, for each replica we use a redundant parametrization
provided by a neural network. Neural networks are a class of
algorithms designed in order to extract information from noisy or
incomplete data, without having to make assumptions on the underlying
law which is obeyed by the data. The only assumption is a certain
degree of smoothness of the function which describes the data. Neural
networks are non-linear functions defined recursively as layers of
nodes which receive inputs from others nodes, and give an output which
is fed to nodes of the next layer.  As an example, in a simple case
with one input $\xi_1^{(1)}$, two hidden neurons and one output
$\xi_1^{(3)}$, (1-2-1), we have
\bea
\xi_{1}^{(3)}=
\frac{1}{1+e^{\theta_1^{(3)}-\xuno-\xdue}}
\nonumber
\eea
where $\omega^{(i)}$ (weights) and $\theta^{(i)}$ (thresholds) are the
parameters of the $i$-th layer. Data are fitted evolving the PDF from
the initial scale to the scale of data, and comparing a physical
observable thus computed to the data in order to tune the best-fit
form parameters of the input PDF, now given by a neural network.

When a large number of parameters is fitted and when correlations
between them are large, as it is the case with a redundant
parametrization, the usual minimization techinques are not optimal. We
have thus implemented a Genetic Algorithm technique
\cite{Rojo:2004iq}, based on mutation and selection of copies of a
given parameters set. The main advantage of the genetic minimization
is that it works on a population of solutions, rather than tracing the
progress of one point through parameter space. Thus, many regions of
parameter space are explored simultaneously, thereby lowering the
possibility of getting trapped in local minima. 

The feature of neural networks which solves the problems of the bias
imposed by a choice of functional form is the fact that the
minimization of a very redundant neural network can be performed, and
stopped when a suitable criterium is met, but before the minimum
$\chi^2$ is reached. This is to be contrasted with standard fits,
where one reaches the lowest $\chi^2$ compatible with the given
functional form, and eventually if one increases the size of the
fitting function no stable fit can be obtained.  The stopping
criterium is the following.  For each replica we separate randomly
data into two sets: one of them is fitted and the other one is
predicted. Since both sets represents the same physical quantity, the
accuracy on both must be same. When $\chi^2$ on the trained set goes
on improving, while on the predicted one starts growing or
oscillating, we stop the minimization. From then on the fit would only
be learning the noise of the fitted set.

\section{Results}

\begin{figure}[t]
\begin{center}
\includegraphics[scale=0.80]{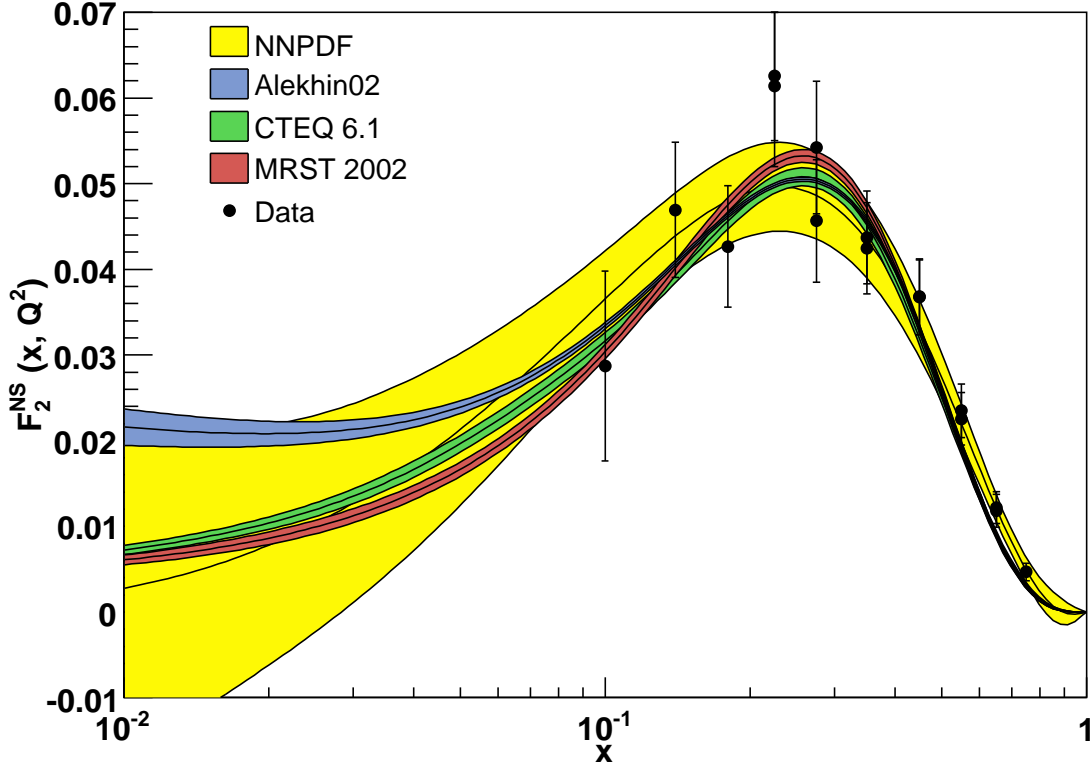}
\caption{NNPDF best-fit of the non-singlet structure function
at $31~{\rm GeV}^{2}< Q^{2} <33~{\rm GeV}^{2}$
compared to other PDF sets. Our 1-$\sigma$ error band has been
evaluated with 1000 replicas; the initial scale PDF is a (2-5-3-1) 
neural network. \label{f2ns}}
\end{center}
\end{figure}

A full determination of $F_2^{NS} (x, Q^2)=F_2^{p} (x, Q^2)-F_2^{d} (x, Q^2)$ 
has been performed with this method. We show our best-fit
for a given $Q^2$ bin compared to the results
obtained by different PDF sets
\cite{Alekhin:2002fv,Pumplin:2002vw,Martin:2002aw}. 
From Fig. \ref{f2ns} we can see that the experimental points
have large errors due to the fact we are taking a difference between
two measurements, while the predictions given by the PDF sets have
smaller errors, since they combine different measurements for the same
points, and since due to evolution points with larger/smaller $x$ and
$Q^2$ carry the same amount of information of the ones shown in the
plot. In the extrapolation region at
small $x$ the different behavior between the Alekhin's fit and the
other sets is due to the fact that Alekhin does not assume any
Regge-like constraint. If no assumption is made on the shape of the
PDF, we obtain a result which agrees both with data and with the other
sets within errors, describes better the large $x$ range
and predicts a larger uncertainty where there is no data. 
One may argue that our error band is wider than the other PDF sets
since these are obtained by fitting much more data than us. However,
since this flavour combination is only constrained by the non-singlet
data which we also use, the small error bands obtained with the
standard approach are more likely due to the way errors are propagated
and to choice of a particular functional form for the initial PDF.
Other examples of the underestimation of errors have been shown in
\cite{Abbate:2005ct}, where a larger error for the Gottfried sum rule
is obtained once the propagation of errors is performed without 
a linearized approximation, and in \cite{Forte:2002us}, 
where a larger error on 
$\alpha_s$ is obtained once no assumption on the PDF shape is made
to fit data.

Further details on our techique and more results
will be given in a forthcoming paper.

\end{document}